\documentclass{iau}
\usepackage{natbib}
\usepackage{graphicx}
\usepackage{float}
\usepackage{url}

\newcommand{\hmpc}{\,$h^{-1}$\,Mpc}

\newcommand{\origami}{{\scshape origami}}

\newcommand{\bq}{\mathbf{q}}
\newcommand{\bx}{\mathbf{x}}

\newcommand{\bpsi}{\mathbf{\Psi}}

\title[An Origami Approximation] 
{An Origami Approximation to the\\
Cosmic Web}

\author[Mark C.\ Neyrinck]   
{Mark C.\ Neyrinck$^1$
\affiliation{$^1$Department of Physics and Astronomy, The Johns Hopkins University, Baltimore, MD 21211 \\ email: {\tt neyrinck@pha.jhu.edu}}}

\pubyear{2014}
\volume{308}  
\jname{The Zel'dovich Universe}
\editors{R. van de Weygaert, S. Shandarin, E. Saar, Jaan Einasto, eds.}
\begin{document}

\maketitle

\begin{abstract}
The powerful Lagrangian view of structure formation was essentially introduced to cosmology by Zel'dovich.  In the current cosmological paradigm, a dark-matter-sheet 3D manifold, inhabiting 6D position-velocity phase space, was flat (with vanishing velocity) at the big bang.  Afterward, gravity stretched and bunched the sheet together in different places, forming a cosmic web when projected to the position coordinates.

Here, I explain some properties of an origami approximation, in which the sheet does not stretch or contract (an assumption that is false in general), but is allowed to fold. Even without stretching, the sheet can form an idealized cosmic web, with convex polyhedral voids separated by straight walls and filaments, joined by convex polyhedral nodes. The nodes form in `polygonal' or `polyhedral' collapse, somewhat like spherical/ellipsoidal collapse, except incorporating simultaneous filament and wall formation. The origami approximation allows phase-space geometries of nodes, filaments, and walls to be more easily understood, and may aid in understanding spin correlations between nearby galaxies.  This contribution explores kinematic origami-approximation models giving velocity fields for the first time.
\end{abstract}

\begin{figure}[ht]
  \begin{minipage}{0.42\linewidth}
    \begin{center}
     \includegraphics[width=\columnwidth]{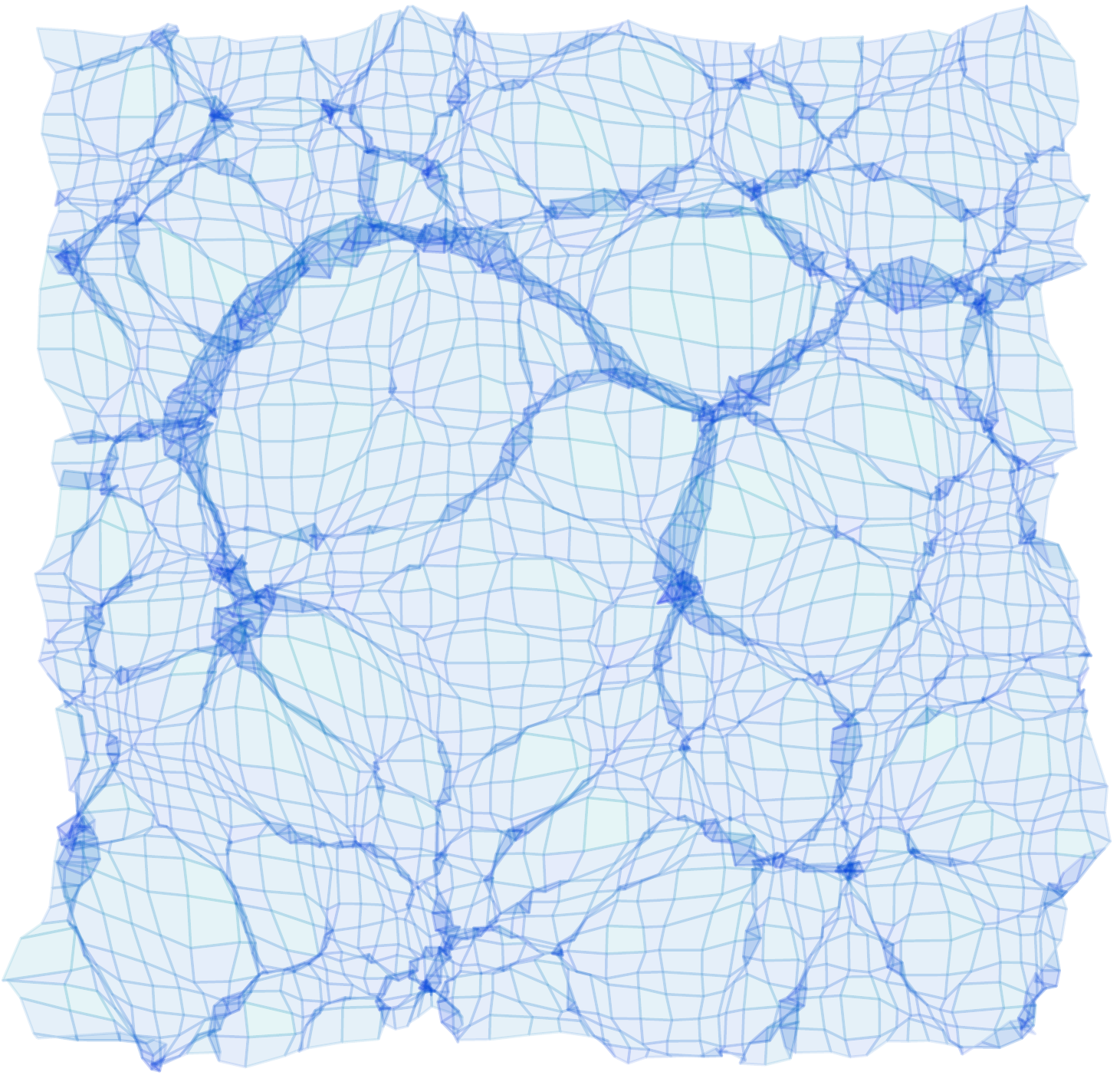}
    \end{center}  
  \end{minipage}
  \begin{minipage}{0.58\linewidth} 
    \begin{center}
      \leavevmode
    \end{center}
  \caption{A dark-matter sheet in a 200\hmpc\ 2D patch, distorted and folded according to the Zel'dovich approximation. The darkness of the color at each position gives the number of streams there. Initially, all vertices were nearly on a regular lattice. Since then, gravity has distorted the mesh, causing regions with a bit more matter than average to accumulate more matter around them. Nodes correspond to galaxies or clusters of galaxies.}
    	\label{fig:fyou}
    \end{minipage}
\end{figure}

In the current structure-formation paradigm, walls, filaments, and cosmic-web nodes in the Universe form somewhat like the origami-folding of a sheet. In paper origami, a 2D non-stretchy sheet is folded in three dimensions. In cosmological structure-formation, the sheet is a stretchable 3D manifold, which folds up in symplectic 6D position-velocity phase space. As in paper origami, the sheet is continuous and cannot tear, and also cannot pass through itself in 6D. This concept is essentially a Lagrangian fluid-dynamics framework (following mass elements, not using a fixed spatial coordinate system). If the matter were not collisionless (dark), Lagrangian patches could not pass through each other in position space to form folded structures. 

Fig.\ \ref{fig:fyou} shows an example cosmic web folded from a collisionless dark-matter sheet, whose vertices have been displaced according to the Zel'dovich approximation. This is a projection of a 2D dark-matter sheet, residing in 4D phase space, down to position space.

In catastrophe theory, singularities (caustics, or folds) that can occur in this sheet can be formally classified \citep{ArnoldEtAl1982,HiddingEtal2014}. However, these descriptions apply only to the local vicinity of caustics. Here, we aim for a description of collections of caustics, that occur at `nodes' that approximate the outer caustic geometry of haloes.

The origami approximation imposes the following assumptions on the displacement field $\bpsi(\bq)$, giving the comoving displacement of a mass element from the initial to final conditions: $\bpsi(\bq)\equiv \bx(\bq)-\bq$, where $\bq$ is a Lagrangian position coordinate, and $\bx(\bq)$ is the position at some later time.  The first two assumptions are thought to hold rather well in reality, but the third is unrealistically strong in general. See \citet{Neyrinck2014origami} for a complementary discussion of this model, aimed at origami researchers.
\begin{enumerate}
\item After applying a sufficiently large smoothing filter to the displacement field, all of space is single-stream \citep[{\it void};\,][]{FalckEtal2012,FalckNeyrinck2014,ShandarinMedvedev2014}. Also, at any resolution being considered, void regions exist.
\item  In void regions, $\bpsi(\bq)$ is irrotational. This is reasonable, since any initial vorticity decays with the expansion of the Universe, and since gravity is a potential force. However, in multi-stream regions (where $\bx(\bq)$ is many-to-one), the flow, averaged among streams, often carries vorticity \citep[e.g.][]{PichonBernardeau1999}.
\item The sheet does not inhomogeneously stretch, i.e.\ $\bx(\bq)$ is piecewise-isometric, or $\mathbf{\nabla}_{\rm Lagrangian}\cdot\bpsi=0$, except at caustics, where it is undefined.  This is the only assumption that is manifestly broken in reality; however, as \citet{HahnEtal2014} have found in warm-dark-matter simulations, the velocity divergence is perhaps surprisingly uniform. It remains largely positive even in filaments, except where it is undefined at caustics.  Despite its inaccuracy, we expect that the piecewise-isometric assumption still allows some correct conclusions to be drawn.
\end{enumerate}

\setcounter{section}{1}
\section{Collapsed structures in 1D and 2D}

In 1D, collapsed structures are delineated by their outer caustics; for example, the canonical phase-space spiral in 2D phase space. From here on, we focus on the outer caustics of structures, by looking at `simple' collapsed regions, i.e. without inner phase-space windings.

In 2D, the simplest collapsed structure is a {\it filament}, an extrusion of a 2D node. With the piecewise-isometry assumption, a filament must consist of straight caustics \citep{DemaineOrourke2008}. Furthermore, the caustics must be parallel, since two reflections produced by non-parallel caustics would cause neighboring voids to be rotated with respect to each other. Here I focus on the outer caustics, and ignore substructure within collapsed regions, but a filament in 2D can form with multiple parallel caustics too; for example, it can form an intricate phase-space spiral in cross-section. 

`Polygonal collapse' occurs at a polygonal {\it node} (intersection of filaments). At a vertex of the polygon, i.e.\ intersection of caustics, Kawasaki's theorem \citep{Kawasaki1989creases} (both alternating sums of vertex angles add to $180^\circ$) implies that all angles between caustics are acute. Thus, nodes are convex. Each vertex also joins an even number $\ge 4$ of caustics. Thus, nodes cannot form in isolation; they must form in conjunction with e.g.\ filaments. Kawasaki's theorem, with some simple geometry, also implies that angles at which filaments come off of the node's edges must equal each other (e.g.\ \cite{Kawasaki1997}). From Lagrangian to Eulerian space, polygonal collapse is a rotation of a convex polygon, producing filaments simultaneously.

Note that circular (spherical) collapse is impossible without stretching the dark-matter sheet, because caustics must be straight lines. Circular (spherical) collapse can be seen as one special case of the collapse of a region, with isotropy around the node, and in which the sheet stretches substantially.  Polygonal (polyhedral) collapse can be seen as another special case, with anisotropy, but no stretching.

\begin{figure}
  \begin{center}
    \includegraphics[width=0.85\columnwidth]{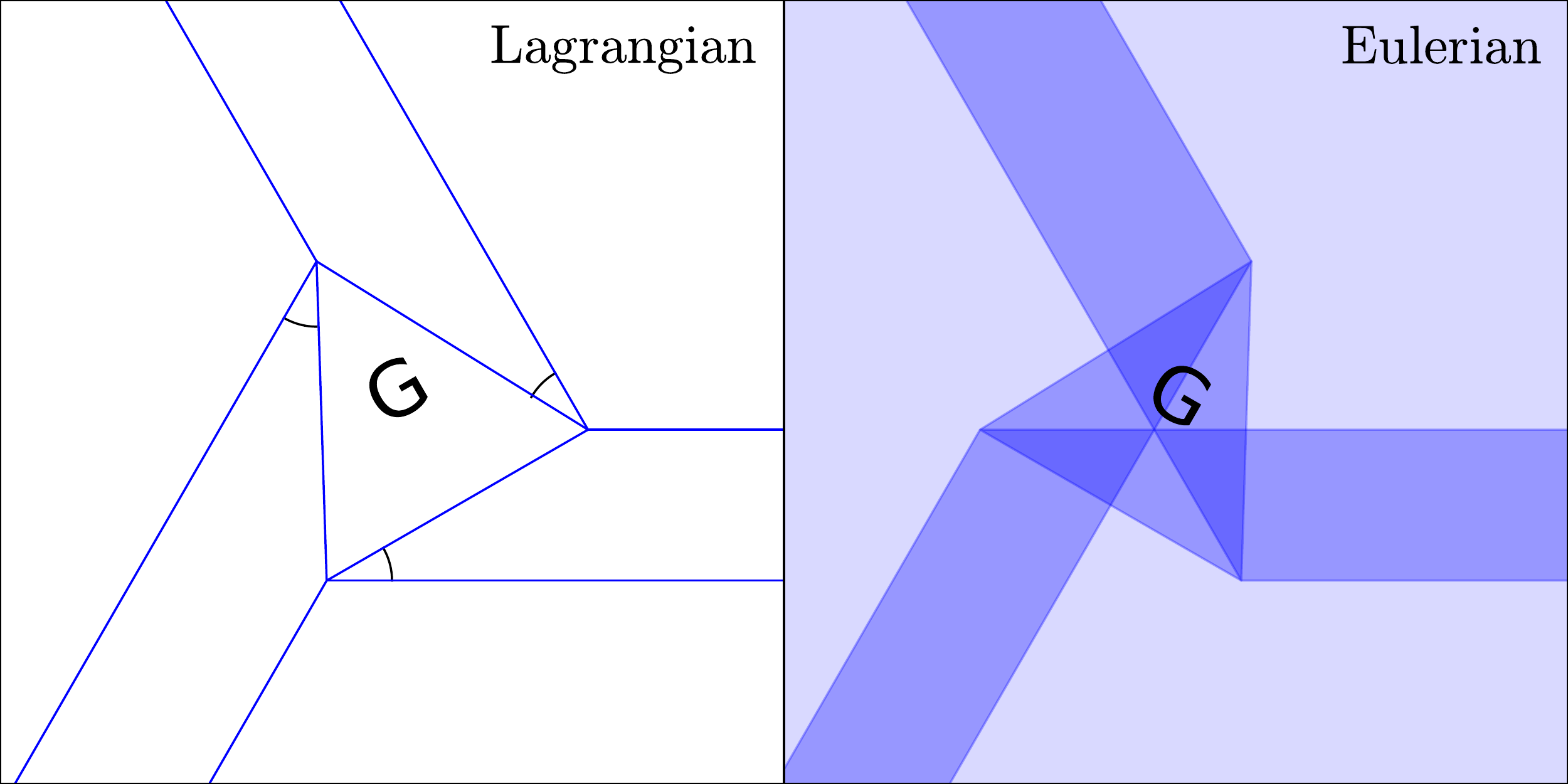}\\
        \textbf{Top:} $\alpha=30^\circ$. \textbf{Bottom:} $\alpha=90^\circ$.\\
    \includegraphics[width=0.85\columnwidth]{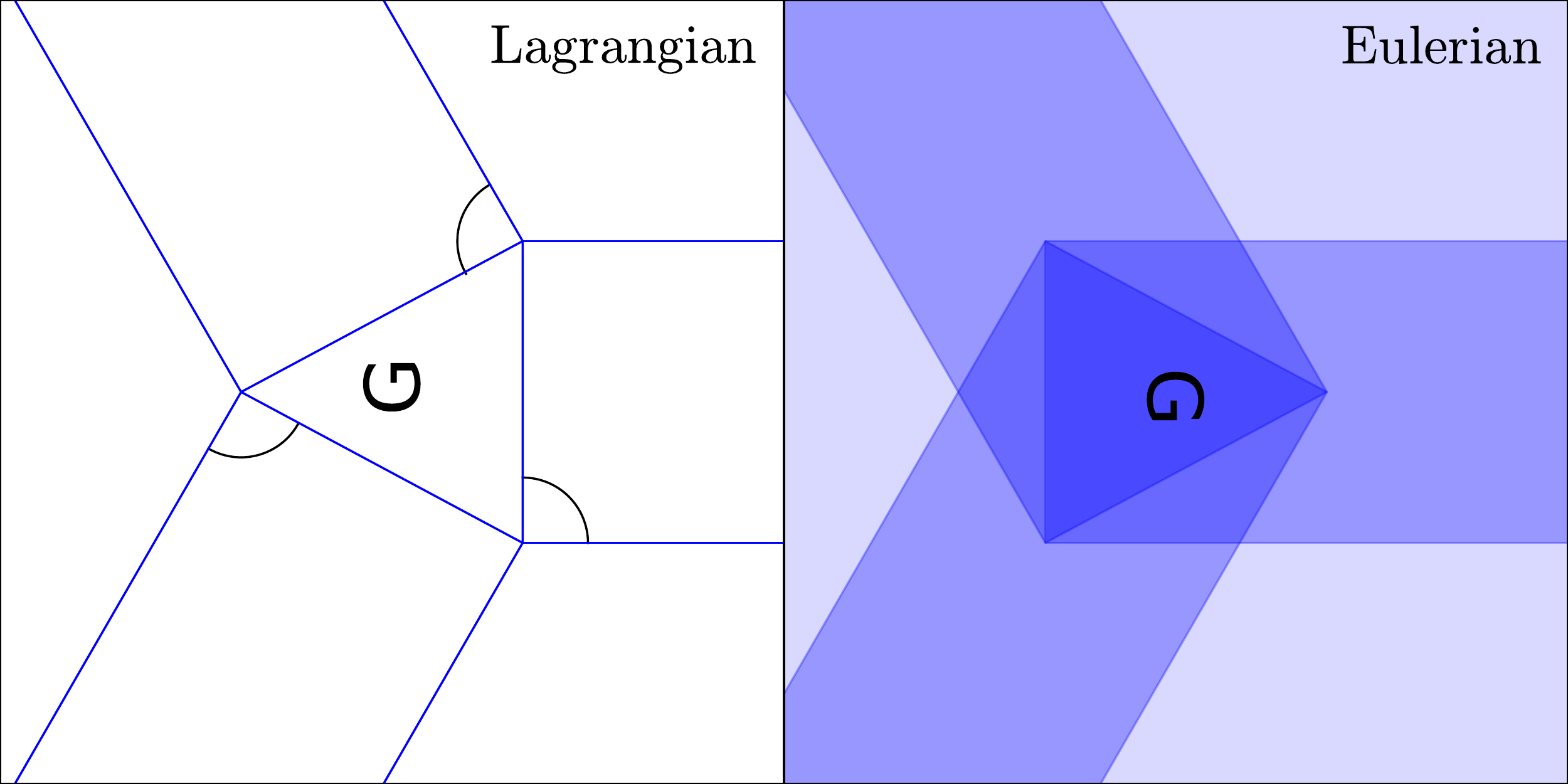}\\
    \end{center}
    \caption{Triangular-collapse models, with different rotation angles $\alpha$. The bottom panels show irrotational collapse, the closest analog to spherical collapse if stretching of the dark-matter sheet is allowed.}
  \label{fig:showdens}
\end{figure}

Fig.\ \ref{fig:showdens} shows two examples of polygonal collapse. At left are the various regions and their caustics in Lagrangian space; at right they appear in Eulerian space. In both, the collapsing polygon is an equilateral triangle, but they differ in the angle $\alpha$ (indicated by arcs). From Lagrangian to Eulerian space, the triangle rotates by $2\alpha$. The bottom panels show irrotational triangular collapse, which can be considered a parity inversion of all elements, since a rotation by 180$^\circ$ is a 2D reflection. Irrotational ($\alpha=90^\circ$) collapse is the closest analog to spherical collapse in the origami approximation.

For a triangular node forming along with filaments, regions can exist with 1 (void), 3 (filament), 5, and 7 streams. In Lagrangian space, the node could be defined simply as the central polygon, but in Eulerian space, it is ambiguous how to delineate the node.  In 2D, the \origami\ algorithm \citep{FalckEtal2012} (not to be confused with the present origami approximation) defines voids, filaments and nodes to have undergone collapse along 0, 1, and 2 orthogonal axes.  For these particular models, \origami\ classifies 1, 3, and $(\ge5)$-stream models as voids, filaments, and nodes, respectively. According to this definition, the central triangle is not classified as a node automatically, unless it rotates by $>90^\circ$ (with $\alpha>45^\circ$). 3-stream regions of the triangle exist in Eulerian space, classified as filament instead of node. At bottom, the entire hexagon is classified as a node. The `outer caustic' of the node is propeller-shaped at top, and hexagonal at bottom.

The origami approximation also imposes restrictions on the topology of cosmic-web components. Since filaments must consist of straight, parallel lines, they are either infinite, or terminate at nodes. Thus, all voids are either infinite, or are convex polygons, forming a tessellation \citep[e.g.][]{Gjerde2008}. This idealized property does not seem to hold in reality; we \citep{FalckNeyrinck2014} found that voids in an $N$-body simulation generally percolate through space, not forming such convex structures. These properties generalize to 3D, as well.

\section{Kinematic models}
Velocity fields can be defined in a kinematic model, with time-varying model parameters. A polygonal-collapse model with fixed shape has two parameters: $s$, the scale of the model (e.g.\ the side length of the triangle); and $\alpha$, the angle. Velocities may be defined from $s(t)$ and $\alpha(t)$.  Thus far, we have not discussed the requirement that the dark matter sheet cannot cross itself in phase space; this may be tested by checking that at each position, velocities on all streams differ.

\begin{figure}
  \begin{center}
    \includegraphics[width=\columnwidth]{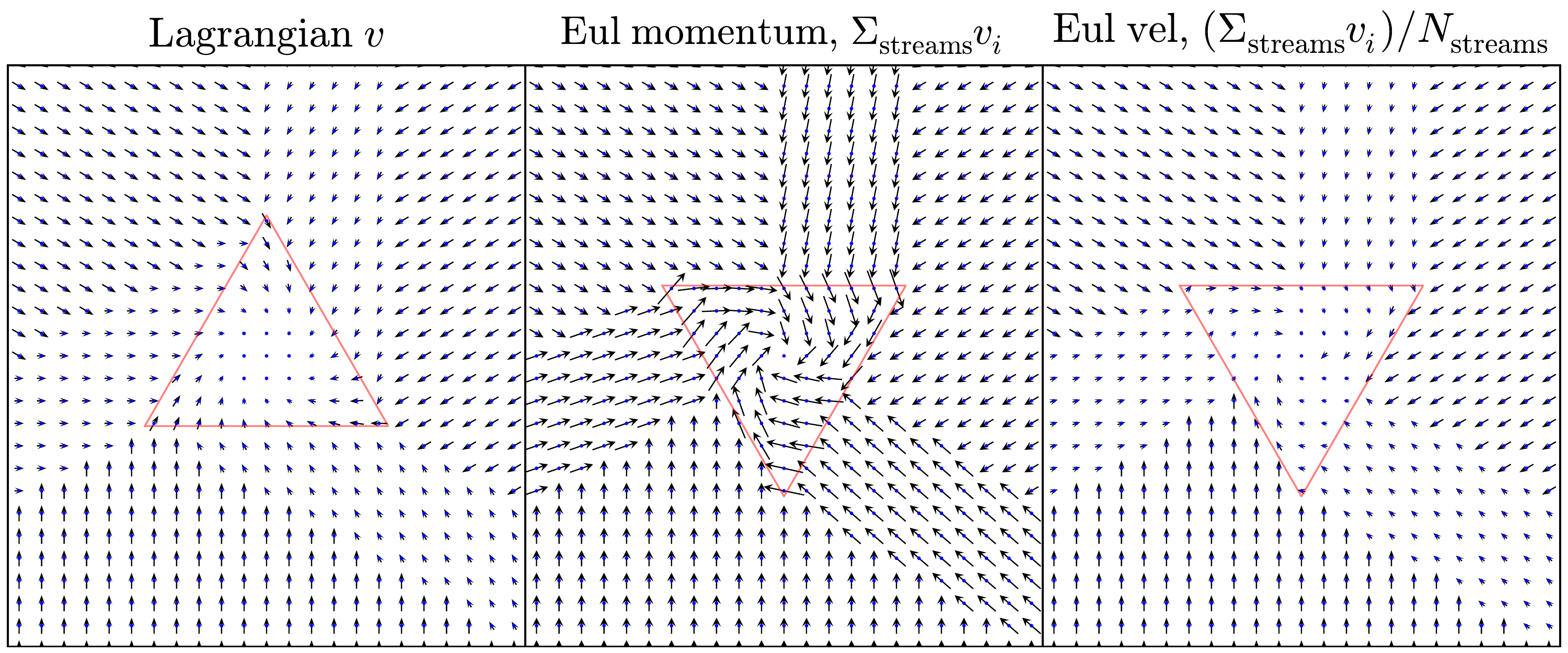}
    \textbf{Top:} $s>0$, $\alpha=30^\circ$, $\dot{\alpha}>0$, $\dot{s}=0$. \textbf{Bottom:} $s>0$, $\alpha=90^\circ$, $\dot{\alpha}=0$, $\dot{s}<0$.
    \includegraphics[width=\columnwidth]{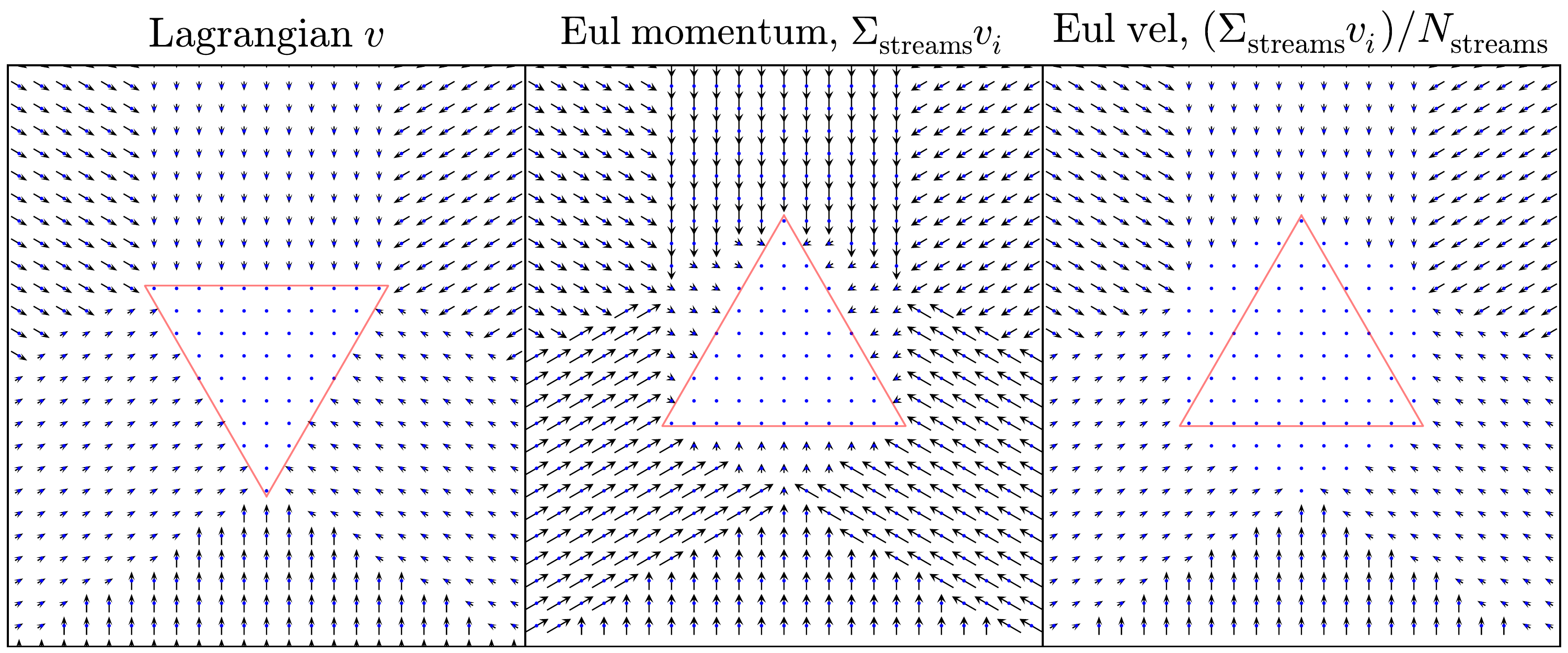}
  \end{center}
  \caption{Kinematic triangular-collapse models, with different values of (the angle) $\alpha$ and (the scale parameter) $s$, and their time-derivatives. {\bf Left}: Lagrangian space. {\bf Middle, Right}: Eulerian space, showing both momentum and mean velocity. {\bf Top}: purely rotational collapse, in which only $s$ increases with time. {\bf Bottom}: irrotational collapse, where only $s$ increases with time. At the current snapshots, these correspond (if tilted by 90$^\circ$) to the models in Fig.\ \ref{fig:showdens}.
  }
  \label{fig:kinematic}
\end{figure}

Fig.\ \ref{fig:kinematic} shows two polygonal-collapse models with different sets of parameters: purely rotational, and purely irrotational models. As expected, there is obvious vorticity in the rotational model in the central node, but not elsewhere. Curiously, the mean velocities (right panels) in multistream regions are lower than in the voids. In the filaments, this is because the velocity is the average over three streams, two of which have rather small components pointing toward the node. The Eulerian momentum, on the other hand, is large, as expected, in filaments.

\begin{figure}
  \begin{minipage}[b]{0.5\linewidth}
     \includegraphics[width=\columnwidth]{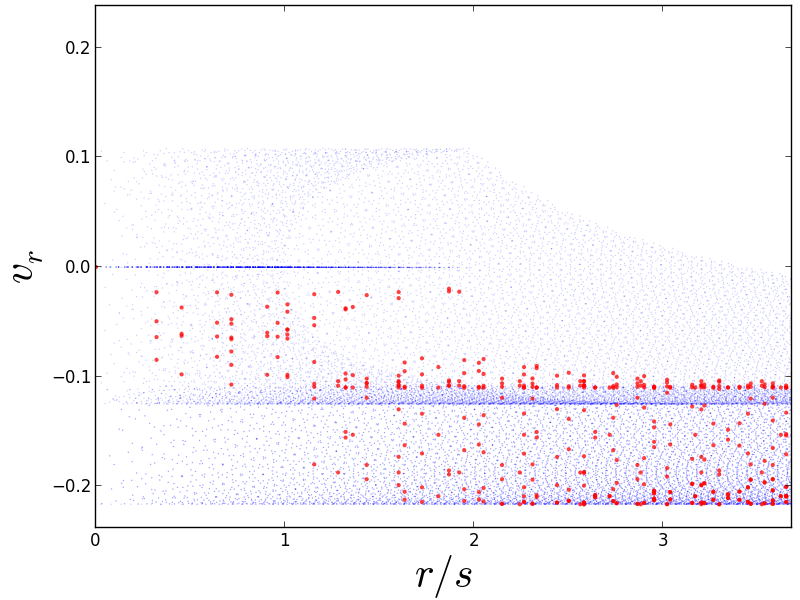}
  \end{minipage}
  \begin{minipage}[b]{0.5\linewidth}
     \includegraphics[width=\columnwidth]{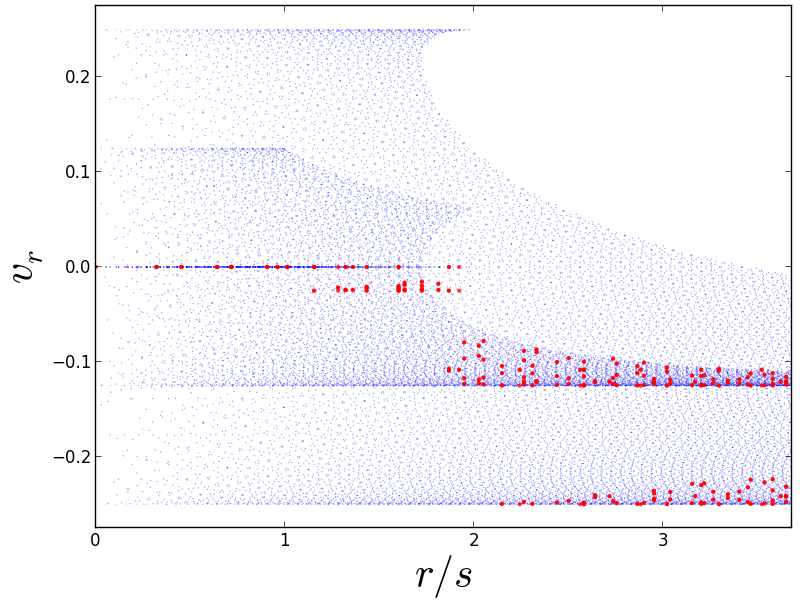}
  \end{minipage}
  \caption{Radial velocities $v_r$ in triangular-collapse models, as a function of radius $r/s$, where $s$ is the incircle radius of triangle (the distance from its center to the nearest edge). Small blue dots show $v_r$ of individual particles, often on different streams, on a fine Lagrangian mesh. Larger, red dots show $v_r$ at the Eulerian positions of arrows in Fig.\ \ref{fig:kinematic}, averaged over streams. {\bf Left}: rotational (corresponding to Fig.\ \ref{fig:kinematic}, top); {\bf Right}: irrotational (corresponding to Fig.\ \ref{fig:kinematic}, bottom).}
    	\label{fig:vr}
\end{figure}

One use of such velocity fields is in practically relating the positions of actual outer caustics to the velocity field around haloes. \citet{DiemerKravtsov2014} have recently proposed a minimum in $v_r(r)$ to delineate outer caustics. Fig.\ \ref{fig:vr} shows $v_r$ explicitly, for the models in Fig.\ \ref{fig:kinematic}. In these plots, blue dots show $v_r$ as a function of Eulerian separation from the node center $r$ for particles on a regular Lagrangian lattice.  Larger, red dots show stream-averaged velocities at Eulerian positions.

For $r\lesssim 3.5s$ (where $s$ is the incircle radius of the triangular node), in both the rotational and irrotational cases, there are particles moving both inward (with negative $v_r$) and outward.  In the rotational case, all Eulerian positions show net inward movement, whereas in the irrotational case, the interior of the node has exactly zero velocity for $r<s$. So, rather surprisingly, a rotational model displays more consistent inward movement than the irrotational model. In this idealized case of a single node, there is no transition to cosmic expansion, which is what would form a minimum in $v_r(r)$, as \citet{DiemerKravtsov2014} find. This behavior might occur if a mean expansion of the sheet is included in a kinematic model. This expansion would be incorporated in a model including several neighboring nodes that remain in place in comoving space, but which grow in mass.

\section{Polyhedral collapse in 3D}
\begin{figure}
  \begin{minipage}{0.5\linewidth}
     \includegraphics[width=\columnwidth]{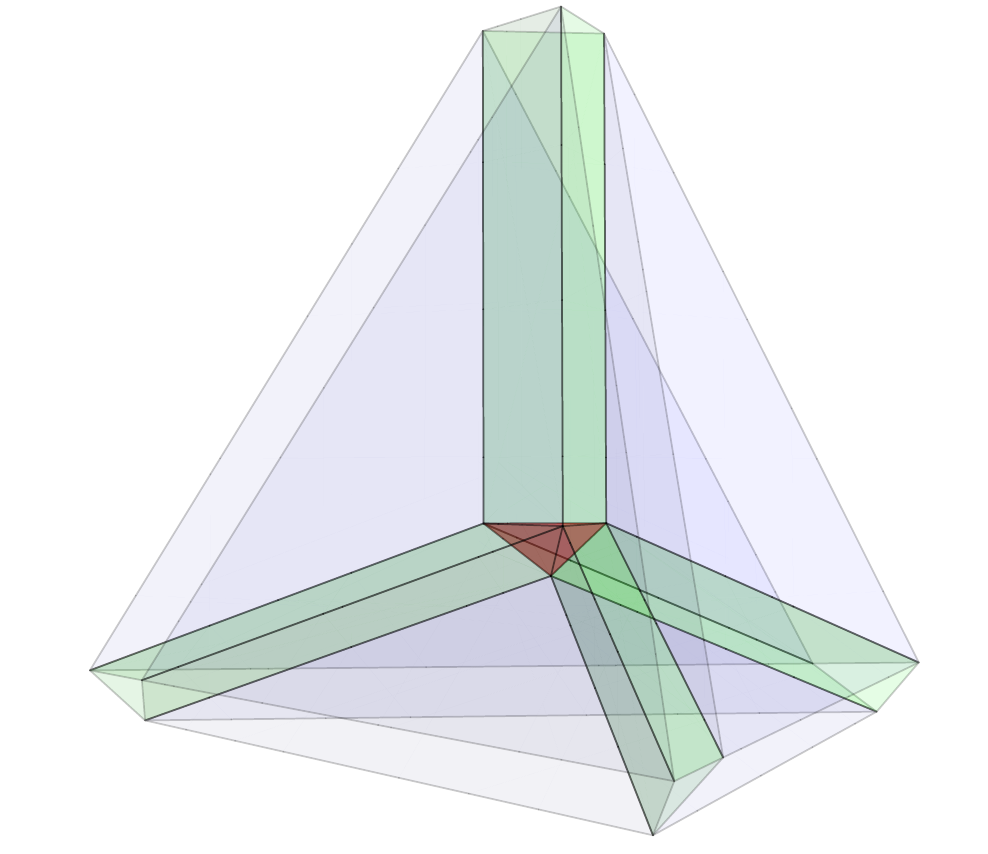}
     \end{minipage}
     \begin{minipage}{0.5\linewidth}
    	\includegraphics[width=\columnwidth]{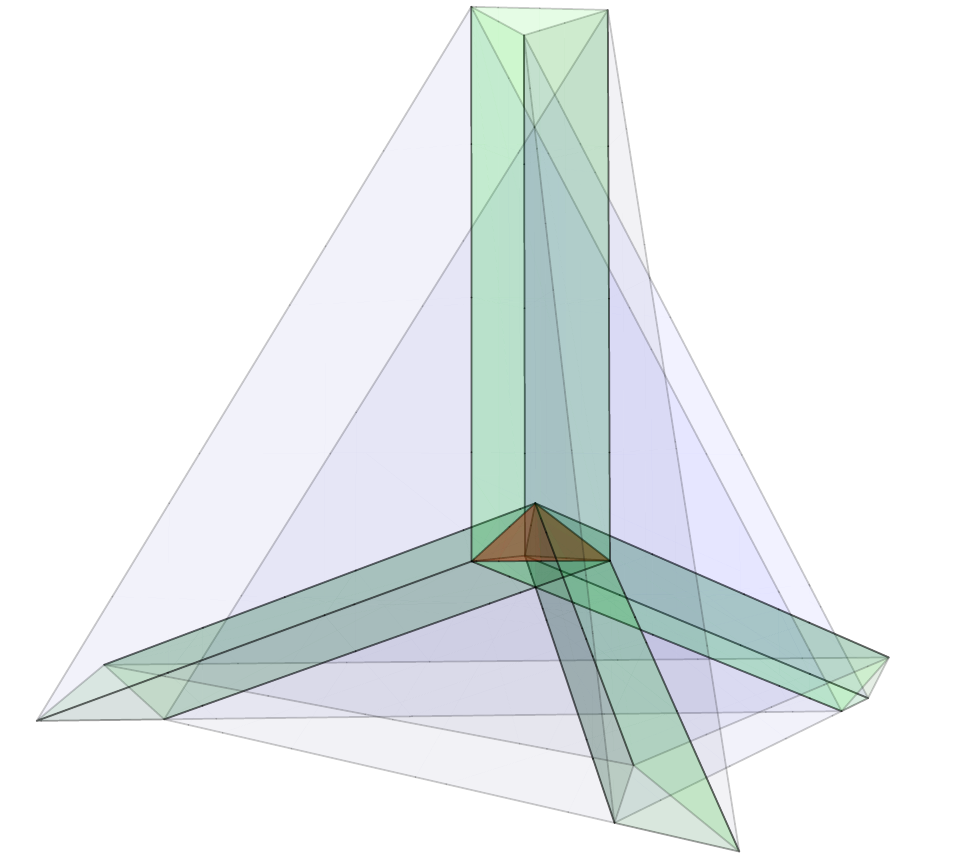}
  \end{minipage}
  \caption{A rotational tetrahedral-collapse model. Filament caustics (green) are triangular tubes, intersecting at the central node. Wall caustics (blue) extend from filament edges through the thin lines drawn between filaments. Node caustics are in red. {\bf Left:} Pre-collapse (Lagrangian). {\bf Right:} Post-collapse (Eulerian). Walls invert along their central planes; the filaments rotate; and the central node both rotates and inverts. The top filament rotates counter-clockwise by $60^\circ$, while the smaller, bottom filaments rotate clockwise by $120^\circ$. See \protect\url{http://skysrv.pha.jhu.edu/~neyrinck/TetCollapse} for an interactive model.}
    	\label{fig:tetcollapse}
\end{figure}

Finally, in Fig.\ \ref{fig:tetcollapse} we make a preliminary foray into 3D, looking at the truly 3D collapse of a tetrahedral node; see \citet{Neyrinck2014origami} for further details of the model. In the origami approximation, although voids and walls are irrotational, filaments can (and in general do) rotate about their axes, correlating the spins of nodes linked by filaments together. The spins of filaments intersecting at nodes are related to each other; in Fig.\ \ref{fig:tetcollapse}, for instance, the top filament rotates counter-clockwise by $60^\circ$, while the smaller, bottom filaments rotate clockwise by $120^\circ$. This can be conceptualized as a set of rods joined by gears in the central node; some of them turn one direction, and others turn oppositely. We will explore the quantitative restrictions this puts on neighboring cosmic-web nodes, comparing to simulations, in future work. 

\bibliographystyle{hapj}
\bibliography{refs}

\end{document}